\documentclass{ws-procs9x6}            
\begin{document}
\title{Exotic Nuclei and Matter in a Chirally Effective Approach}

\author{S. Schramm}

\address{FIAS, Ruth-Moufang-Str. 1,D-60438 Frankfurt am Main, Germany\\
$^*$E-mail: schramm@fias.uni-frankfurt.de}

\begin{abstract}
A relativistic approach to describe nuclear and in general strongly interacting matter is introduced and discussed. Here, not only the nuclear forces but also the masses of the nucleons are generated through meson fields. Within this framework it is possible to calculate properties of finite nuclei at a level of accuracy similar to dedicated relativistic nuclear structure models. Due to the more general approach, a wider range of properties of hadronic states can be investigated. A number of results for heavy and neutron-rich nuclei toward the drip line are presented. 
\end{abstract}

\keywords{chiral model, super heavies, drip lines, exotic nuclei}

\bodymatter

\section{Introduction}

\label{sec:1}
In modern nuclear physics one general area of research tries to push the boundaries of strongly interacting matter towards extremes of temperature, density as well as isospin.
In the case of extreme temperatures the experiments of choice are ultra-relativistic heavy-ion collisions that create a short-lived very hot fireball in the  collision zone with a small net baryon density (i.e. about as many particles as anti-particles). At the other end of conditions neutron stars or more generally, compact stars, feature very large baryon number densities in the core and low temperatures.
Therefore the analysis of neutron star observations can yield indirect hints to the properties of dense strongly interacting matter. However, this is not the only difference with respect to heavy-ion collisions.
As the name indicates, neutron stars are predominantly made up of neutrons, if one neglects possible hyperonic and quark phases. Therefore the isospin density of this system is very high, compared to heavy-ion collisions. Here this discussion connects to exotic nuclei with large neutron numbers exploring the range up to the neutron drip line. For these systems the hadronic matter is sampled at around saturation density, low temperatures and relatively high isospin. Therefore, all these studies complement each other in the understanding of strongly interacting matter under extreme conditions.

While there are a number of theoretical approaches that try to address these various areas separately, in principle a general strong interaction model framework is needed that can describe all the mentioned conditions.
This is the path we follow in this article. We describe such an approach for dense and hot matter that in its full version also includes quark degrees of freedom. However, in this article we concentrate on nuclear properties. 
The interaction parameters of the model are fine-tuned such that a good description of nuclei is possible. Within this framework it is then possible not only to study nuclei and exotic isotopes but also neutron stars and heavy-ion collisions opening up the possibility to establish correlations of observables connecting these very different regimes.

\section{Model Description}

We have followed this path  of developing a general model applicable to a large range of densities and temperatures for a number of years, developing an effective chiral SU(3) hadronic (CMF) model, where the baryonic masses are largely generated by their coupling to scalar fields that attain a vacuum expectation value due to spontaneous symmetry breaking, analogously to simple chiral $\sigma$ models\cite{weinberg}.

The model is described in detail in various publications \cite{Papazoglou:1997uw,Papazoglou:1998vr,deformed}.
Within the CMF approach baryons interact via meson exchange, where the meson fields are approximated by their mean field values.
In consequence, 
the effective baryon masses $m_i^*$ read
\begin{equation}
m_i^* = g_{i\sigma} \sigma + g_{i\zeta} \zeta + g_{i\delta} \delta + \delta m_i ~~.
\label{mass}
\end{equation}
where $i$ labels the various baryons of the lowest baryonic multiplet containing the nucleons as well as the $\Lambda, \Sigma^{+/+/0},$ and $\Xi^{0,-}$ hyperons.
The scalar fields in this equations, $\sigma, \delta, \zeta$, correspond to the scalar quark condensates,
$\sigma \sim <\overline{u}u + \overline{d}d> $,  
$\zeta \sim <\overline{s}s>$, and
$\delta \sim <\overline{u}u - \overline{d}d> $.
As it is a very well established result from QCD, clearly shown in lattice gauge simulations, the vacuum state contains non-vanishing scalar (and isoscalar) condensates, directly related to $\sigma$ and $\zeta$.
In the model approach this is achieved through a  flavor-SU(3) self-interaction that generates non-vanishing scalar fields in the vacuum \cite{deformed}.
A mass term $\delta m_i$ breaks chiral and SU(3) symmetry explicitly. The various couplings contained in the equation result from the SU(3) coupling scheme.
Thus, the non-zero vacuum expectation values of $\sigma$ and $\zeta$ generate the baryonic masses in the vacuum, while the change of the scalar fields at finite density or temperature are responsible for the scalar attraction. 
The non-vanishing vacuum condensates are the result of
the structure of the SU(3)-invariant scalar self-interactions, given by \cite{Papazoglou:1998vr,deformed}
\begin{eqnarray}
&L_{Self}= k_0(\sigma^2+\zeta^2+\delta^2)+k_1(\sigma^2+\zeta^2+\delta^2)^2+k_2\left(\frac{\sigma^4}{2}+\frac{\delta^4}{2}
+3\sigma^2\delta^2+\zeta^4\right)\nonumber\\&
+k_3(\sigma^2-\delta^2)\zeta+k_4\ \ \ln{\frac{(\sigma^2-\delta^2)\zeta}{\sigma_0^2\zeta_0}}~.
\end{eqnarray}
with various coupling constants $k_i$ that have largely to be chosen to reproduce baryon masses \cite{deformed,Schrammsoon}.

The vector fields $\omega, \rho$ ($rho^0$ meson) and $\phi$ are the analogous fields to the $\sigma, \delta, $ and $\zeta$, respectively.  The fields acquire non-zero mean fields in dense matter and generate repulsive forces between baryons via the linear interaction term
\begin{equation}
L_{BV}= - \sum_i \overline{\psi_i} \left[ g_{\omega i} \omega  +  g_{\rho i} \rho  + g_{\phi i} \phi  \right] \psi_i
\end{equation}
summing over the baryons with their fields $\psi_i$. The various couplings are related through SU(6) relations \cite{deformed}.

\begin{figure}
\includegraphics[width=4.5in]{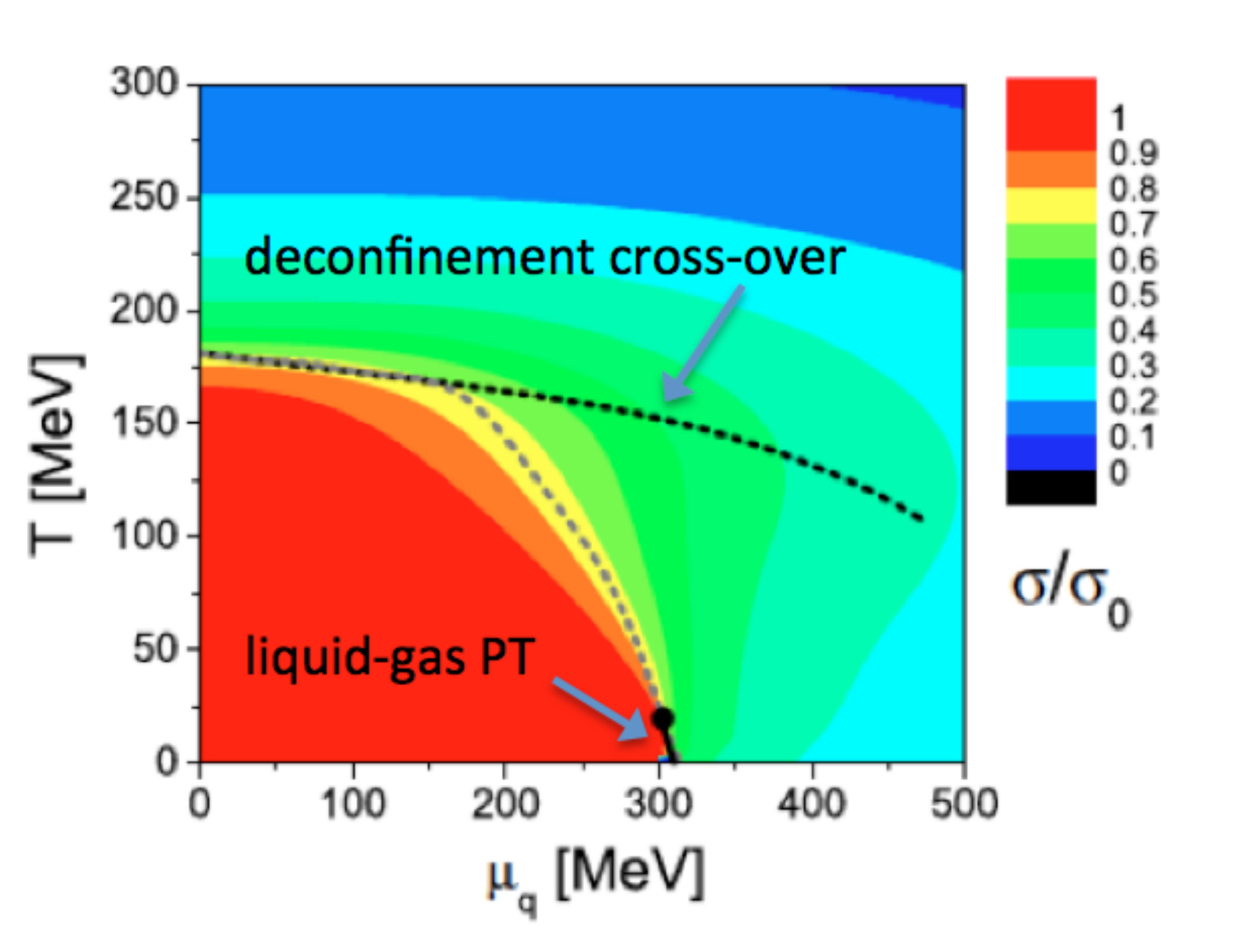}
\caption{Scalar fiel $\sigma$ as function of temperature and baryochemical potential. The shades code the value of the field. Also marked are the first-order liquid-gas phase transition as well as the position of the smooth deconfinement transition.}
\label{fsigma}
\end{figure}

Thus the interplay of the scalar and vector fields produces saturation of nuclear matter. In order to obtain not just reasonable infinite matter results but also a good description of finite nuclei, a careful fine-tuned optimisation of the parameters has to be performed.
To illustrate the usefulness of the approach  it is worth pointing out that the hadronic model can be readily extended to include quarks as well as has been discussed in \cite{Steinheimer:2010ib,Dexheimer:2009hi}
in detail. Both quarks and hadrons couple to the condensate fields and therefore allow for smooth transition from hadronic degrees of freedom at low temperature and/or densities to high densities and temperatures, respectively. This can also be illustrated by calculating the scalar condensate $\sigma$ as function of temperature $T$ and baryon chemical potential $\mu_b$.
The resulting plot is shown in Fig. \ref{fsigma}. One can see the reduction of the condensate at high $T$ and $\mu_b$. The model also shows a first-order liquid gas phase transition.

\begin{figure}
\includegraphics[width=4.5in]{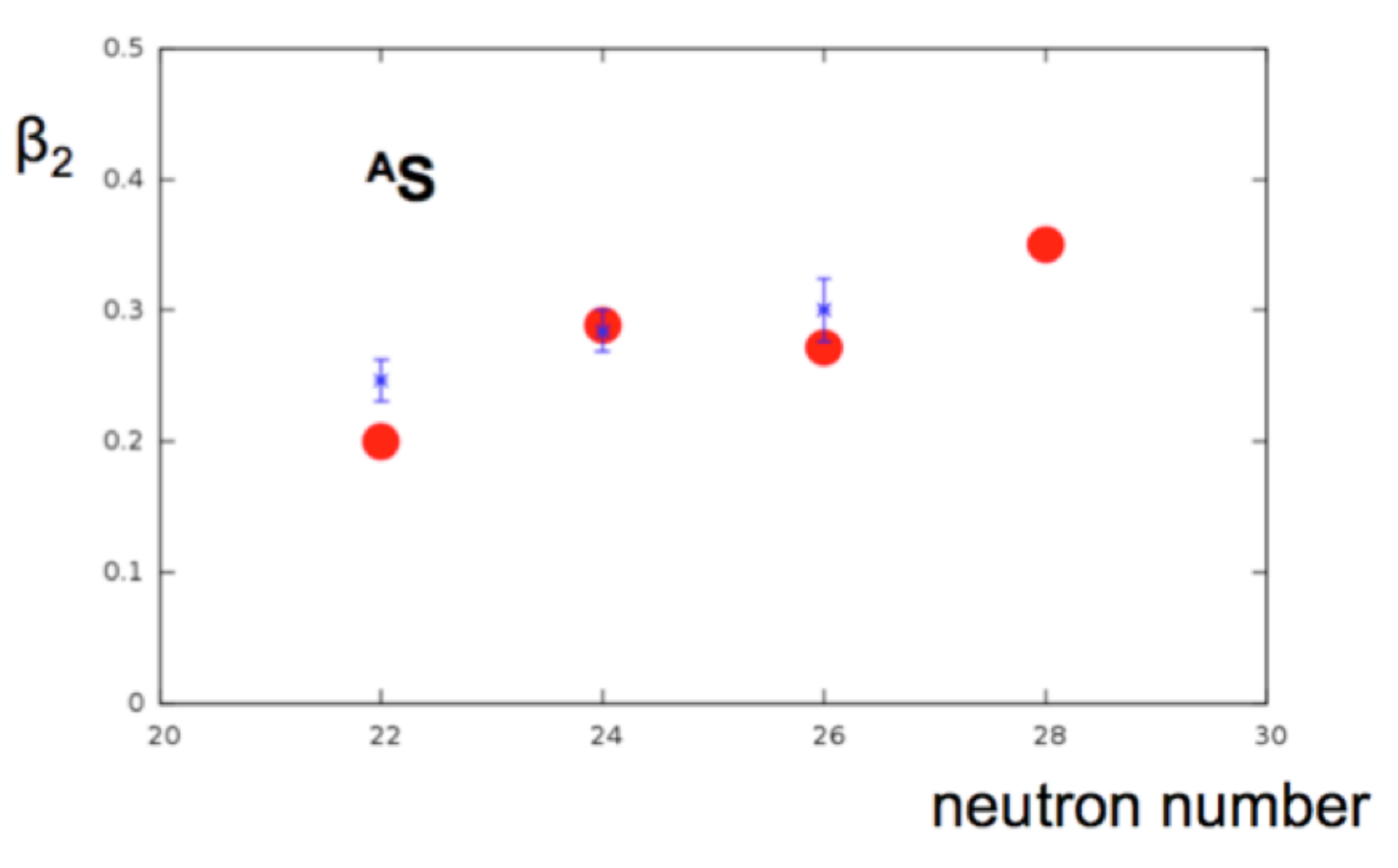}
\caption{Quadrupole deformation of neutron-rich sulfur isotopes. Theoretical and experimental \protect\cite{Scheit:1996zz} values are shown.}
\label{fsulfur}
\end{figure}

\begin{figure}
\includegraphics[width=4.5in]{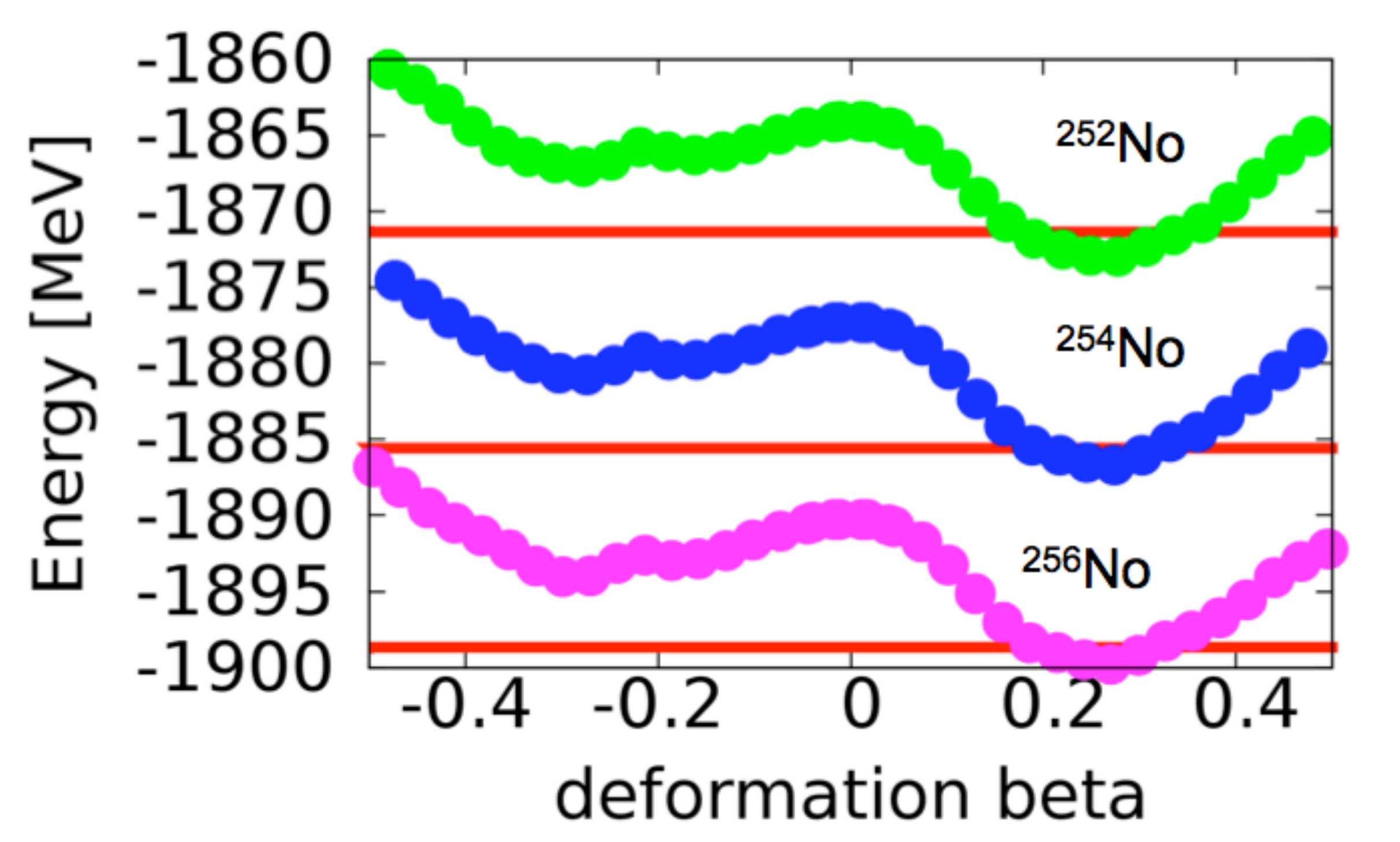}
\caption{Energy of nobelium isotopes as function of quadrupole deformation. The ground states of the nuclei exhibit strong prolate deformation.The horizontal lines mark experimental data.}
\label{fnobel}
\end{figure}

\begin{figure}
\includegraphics[width=4.5in]{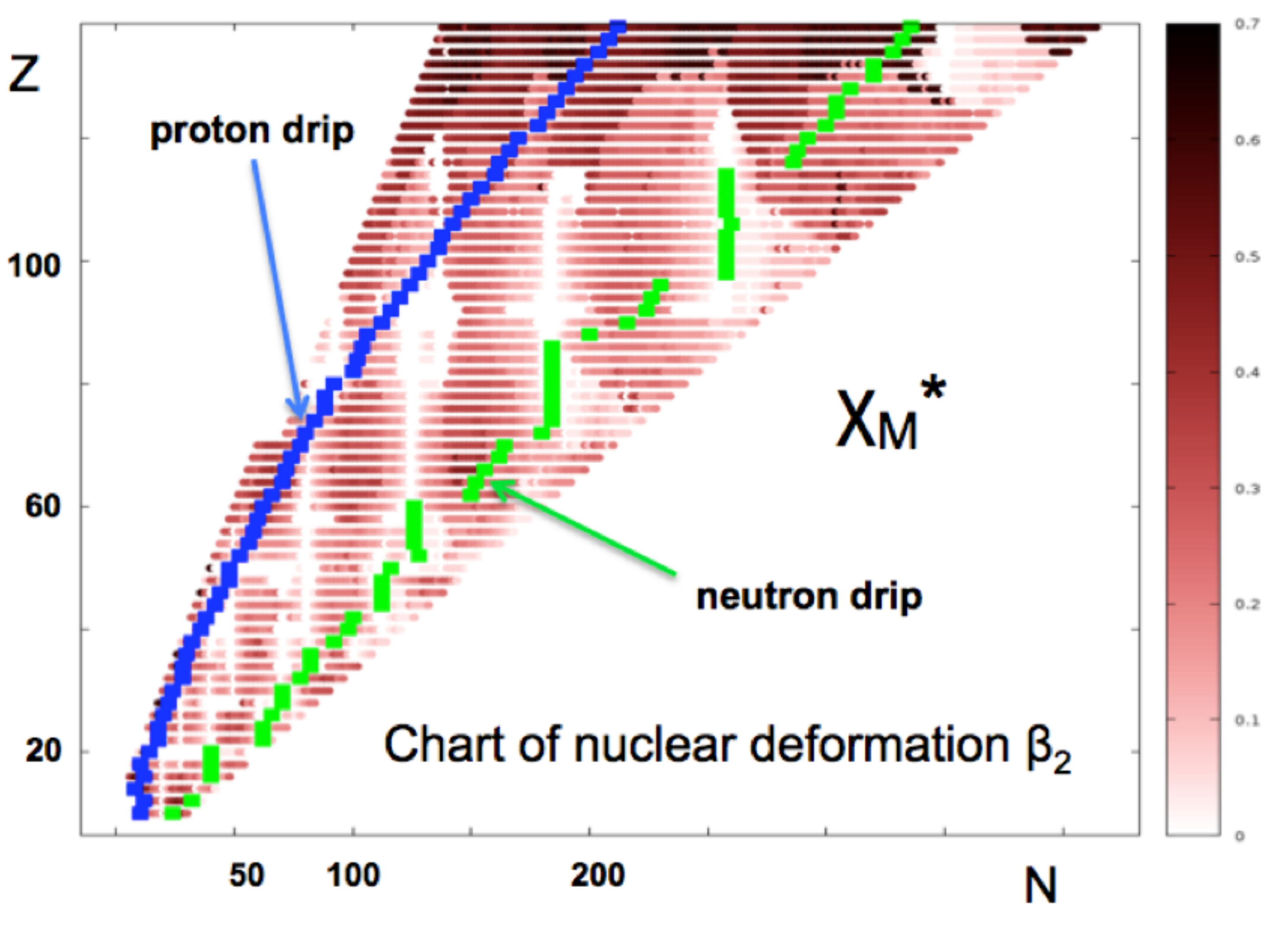}
\caption{Nuclear quadrupole deformation (absolute value) as function of neutron number N and charge number Z. The nucleon drip lines are marked. The darker colours mark stronger deformation.}
\label{fchart}
\end{figure}

\begin{figure}
\includegraphics[width=4.5in]{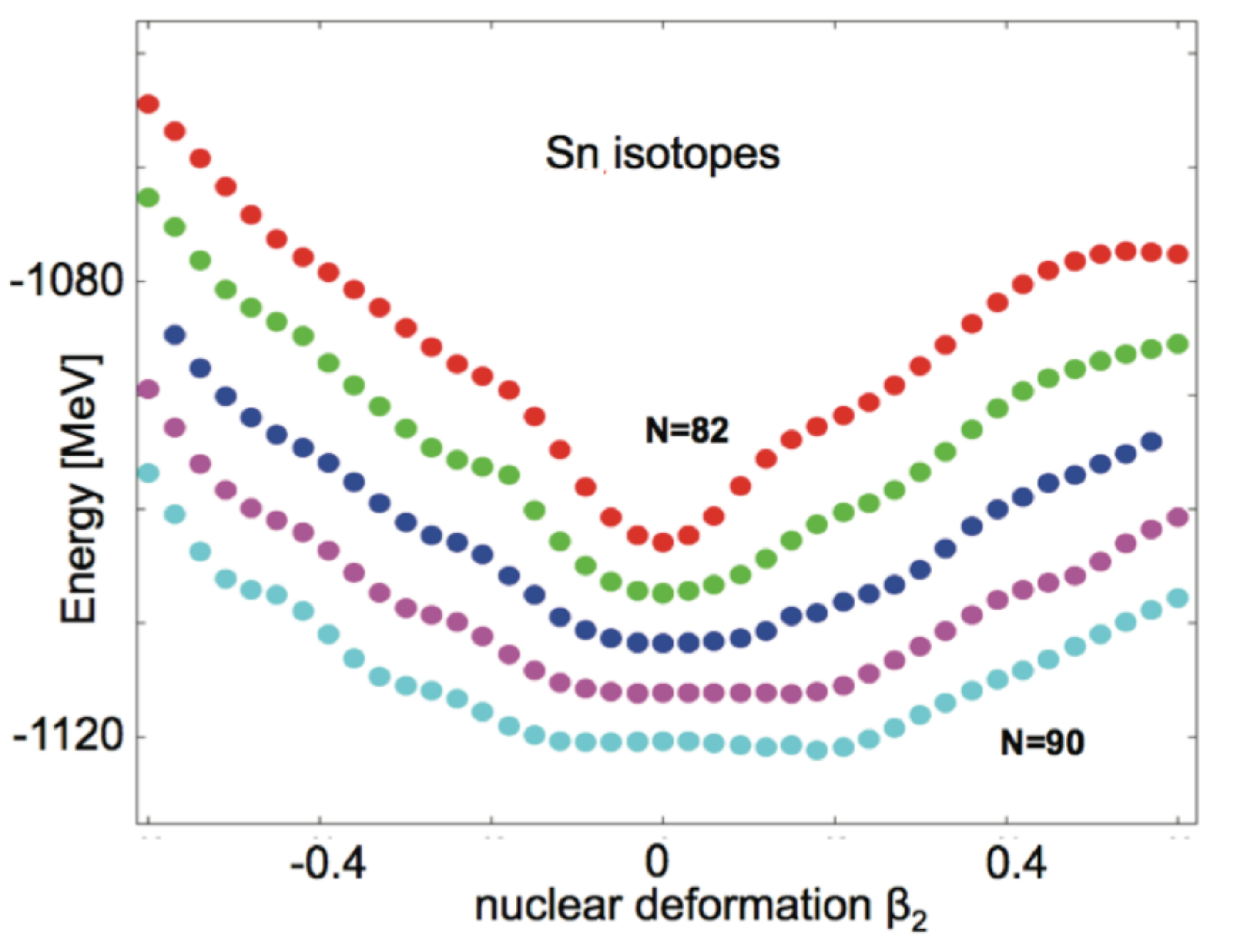}
\caption{Binding energy as function of $\beta_2$ for various very neutron-rich Sn isotopes. The weakening of the effect of the shell closure with increasing neutron number can be clearly observed.}
\label{ftin}
\end{figure}

\begin{figure}
\includegraphics[width=4.5in]{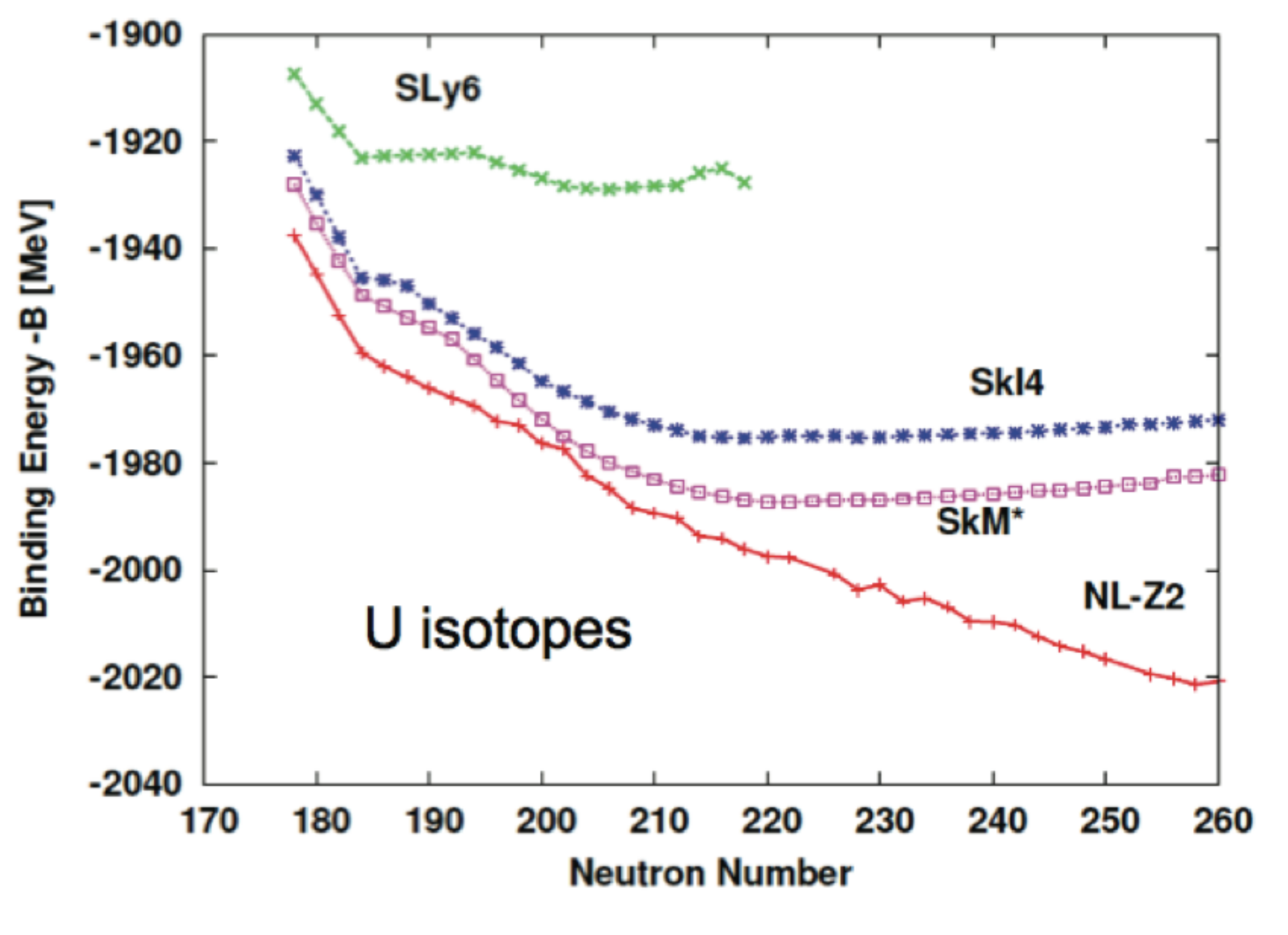}
\caption{Binding energies of uranium isotopes for a range of Skyrme forces and a RMF parameterisation.}
\label{furanium}
\end{figure}

On the purely hadronic side, a very recently improved fit of the coupling constants\cite{Schrammsoon} was used to calculate all even-even nuclei up to the proton and neutron drip lines.
The resulting errors $\epsilon$ in binding energy compared to experimental data \cite{audi} amount to $\epsilon = 0.17 \%$ for nuclei with baryon number larger than or equal to $A = 50$, and $\epsilon = 0.12 \%$ for $A \le 100$., which is a $20 \%$ improvement to earlier results \cite{deformed}. As it is common in the case of mean field approaches, larger systems can be described more accurately.
\\ \\
In addition to a satisfactory description of binding energies of known nuclei a reasonable description of surface effects are an important part of a successful nuclear model. In this context various deformation properties of nuclei were investigated. Fig. \ref{fsulfur} shows the quadrupole deformation $\beta_2$  of sulfur isotopes.
Also shown are measured values for $\beta_2$ \cite{Scheit:1996zz}. The figure shows good agreement of data and model. Furthermore, the prediction is that for increased neutron number ($N=28$) a further rise of the already strong deformation of the sulfur isotope chain is to be expected.
A similar study was performed for heavy nuclei. Fig. \ref{fnobel} presents results for the energy of nobelium isotopes. as function of the quadrupole 
deformation. The horizontal lines mark the experimental binding energy. Measured values for the deformation are $\beta_2(^{252}No) =  0.31 \pm 0.02$ and 
$\beta_2(^{254}No) = 0.32 \pm 0.02$ \cite{Herzberg:2001cj}. As one can see there is quite a good agreement with experiment.
In a survey of particle-stable nuclei we performed a scan of nuclear deformation over the whole range of possible charge and neutron number. The results are
shown in Fig. \ref{fchart}. The color-coded information shows the absolute value of the quadrupole deformation of the nuclei. Also indicated are the proton and neutron drip lines. The results for nuclei beyond the drip line are not reliable, as some nucleons start leaking from the nucleus and the numerical quantities can depend on the box size of the actual simulation. 
\\ \\
Zooming into the very neutron-rich tin isotopes, one can observe the quenching of the shell effects for the $Z=50$ nuclei. Fig. \ref{ftin} shows a pronounced spherical minimum for neutron number $N=82$. Increasing the number a clear softening of this spherical shape can be observed and slowly minima at higher deformation develop.
\\ \\
As an interesting feature of this plot note the development of a strong shell closure for the very large neutron number $N=258$. This rather model-independent shell closure was first discussed in \cite{Schramm:2011ad}. Fig. \ref{furanium} shows the result for Uranium isotopes for a number of different Skyrme forces and the relativistic parameterisation NL-Z2 \cite{Bender:1999}. Here one can observe that already for $Z=92$ one model shows that the maximum binding in the Uranium isotope chain can be obtained for $N=258$. Going to higher $Z$ the other forces will show a similar behaviour pushing the drip line   towards this value.

\begin{figure}
\includegraphics[width=4.5in]{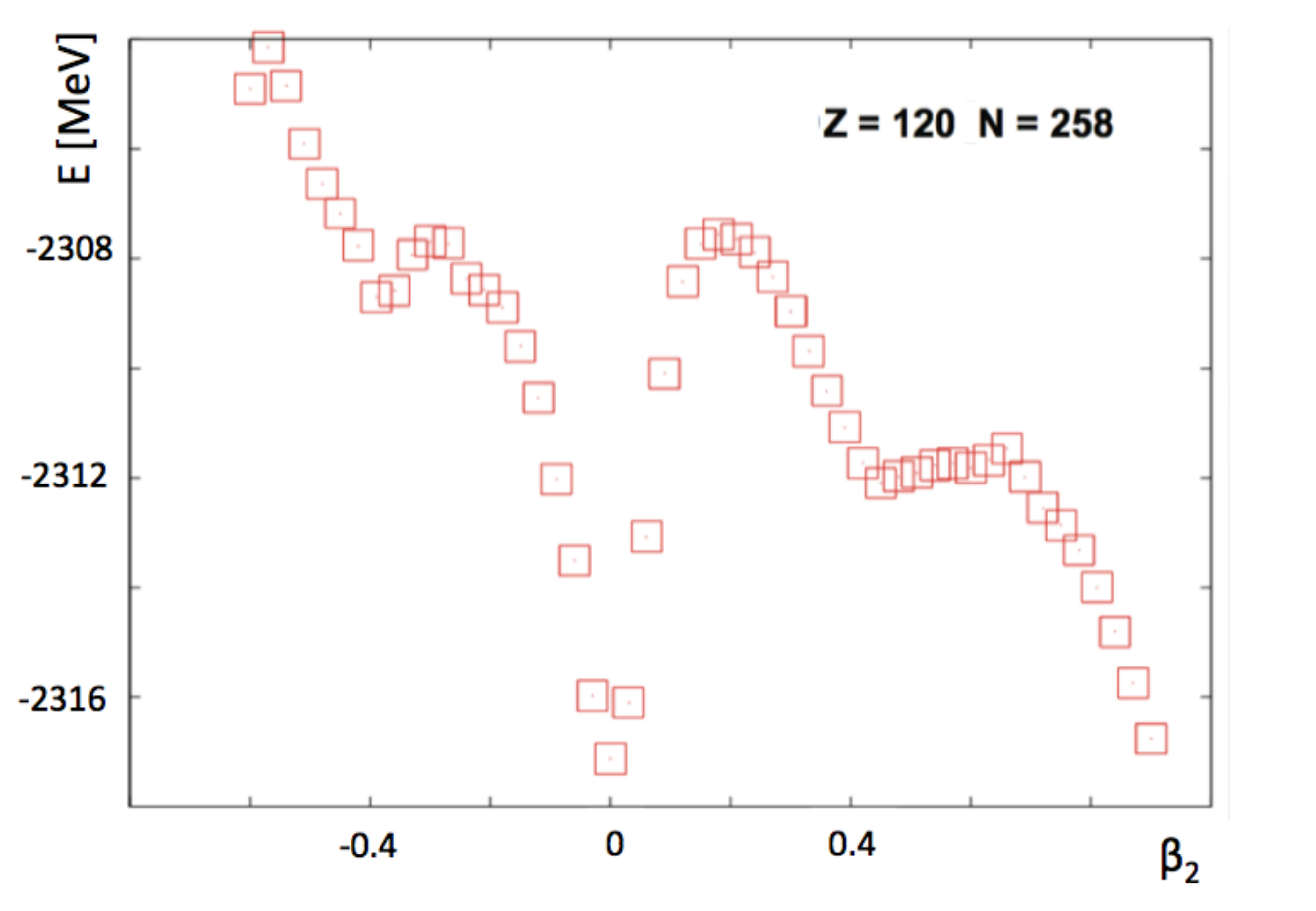}
\caption{Binding energy as function of deformation for the superheavy nucleus $^{378}120$. The curve shows a clear minimum at zero deformation
with a large fission barrier.} 
\label{fsuper}
\end{figure}

In the usual regime of superheavy nuclei around $Z = 114$ or $Z = 120$ with neutron numbers between $N = 170$ and $N = 190$ no clear sign of a spherical nucleus can be observed.
In addition, the fission barriers are quite low of the order of 4 to 5 MeV. In this regime an extended calculation allowing for octupole deformations (that have been neglected in the other numerical studies shown here) shows a significant reduction of the fission barriers as is exemplified in Fig. \ref{z120}. This general behaviour has also been observed in other RMF approaches \cite{Burvenich:2003gz}.
\begin{figure}
\includegraphics[width=4.5in]{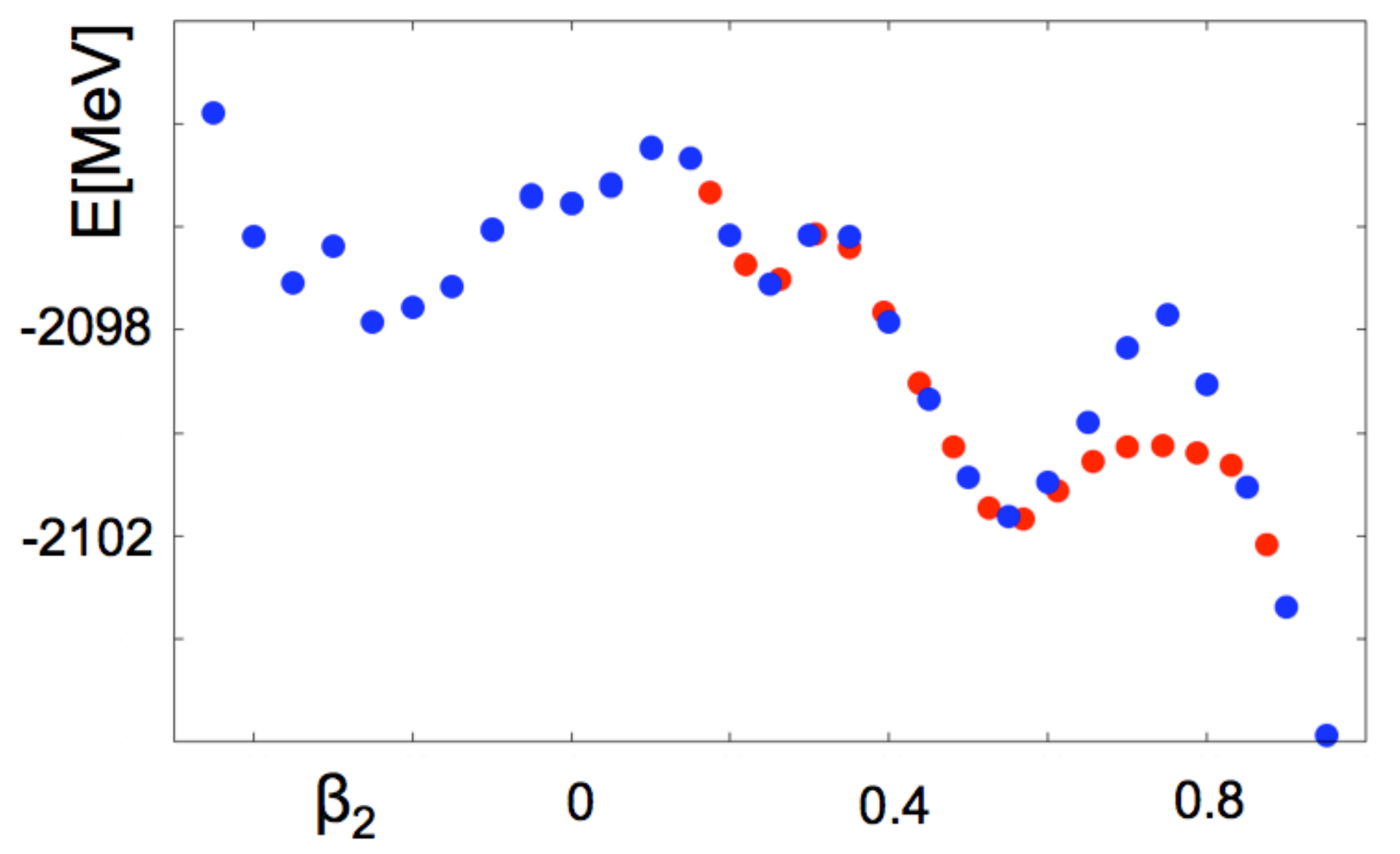}
\caption{Example of a superheavy nucleus with $Z = 120$ and $N = 174$. The (negative) binding energy shows a minimum at large prolate deformation (blue points).
However, including octupole deformations the fission barrier is reduced significantly (red points).}
\label{fz120}
\end{figure}
However, investigating the corresponding extremely neutron-rich isotopes a different picture emerges.
This is demonstrated in Fig. \ref{fsuper}. This graph
shows the deformation-dependent binding energy of a nucleus with charge $Z = 120$ and neutron number $N = 258$.
In contrast to the solutions of superheavy nuclei with smaller neutron numbers and rather small fission barriers, this nucleus exhibits a very
distinct spherical minimum with a fission barrier of about $9$ MeV. A similar result was also found in Skyrme models \cite{Tarasov:2014yoa}.

\section{Conclusion}

We discussed a general flavor-SU(3) approach to study hadronic and quark matter in general, and nuclear structure problems in particular.
For a quantitatively convincing description of nuclear structure this entails a fine-tuning of the parameters \cite{deformed,Schrammsoon}.

The resulting model generates good results for binding energies in comparison with experimental data. Deformation properties of various nuclei also
show agreement with experiment. By extending the calculation to very neutron-rich nuclei the appearance of a strong shell closure at a (very high)
neutron number (N = 258) is quite remarkable. This also leads to extremely stable superheavy nuclei. Whether such states could be produced, for instance in the crust of neutron stars and then ejected by a violent process or a star merger, is still completely open and should be estimated.




\end{document}